\documentclass[amsmath,amssymb]{elsarticle}
\usepackage{graphicx}
\usepackage{dcolumn}
\usepackage{bm}
\usepackage{amsmath}
\usepackage{color}

\newcommand{\eq}[1]{Eq.~(\ref{#1})}    

\begin{document}
\title{The near-extreme density of intraday log-returns}

\author[icu,centrale,bcam]{Mauro Politi}
\ead{mauro@nt.icu.ac.jp}

\author[centrale]{Nicolas Millot}
\ead{nicolas.millot@ecp.fr}

\author[centrale]{Anirban Chakraborti}
\ead{anirban.chakraborti@ecp.fr}

\address[icu]{SSRI \& Department of Economics and Business, International Christian University, 3-10-2 Osawa, Mitaka, Tokyo, 181-8585 Japan}
\address[centrale]{Chaire de Finance Quantitative, Laboratoire de Math\'ematiques Appliqu\'ees aux Syst\`emes, \'Ecole Centrale Paris, 92290 Ch\^atenay-Malabry, France}
\address[bcam]{Basque Center for Applied Mathematics, Bizkaia Technology Park,\\ Building 500, E48160, Derio, Spain}


\begin{abstract}
The extreme event statistics plays a very important role in the theory and practice of time series analysis. The reassembly of classical theoretical results is often undermined by non-stationarity and dependence between increments. Furthermore, the convergence to the limit distributions can be slow, requiring a huge amount of records to obtain significant statistics, and thus limiting its practical applications. Focussing, instead, on the closely related density of ``near-extremes'' -- the distance between a record and the maximal value -- can render the statistical methods to be more suitable in the practical applications and/or validations of models. We apply this recently proposed method in the empirical validation of an adapted financial market model of the intraday market fluctuations.
\end{abstract}


\maketitle

\setcounter{equation}{0}
\section{Introduction}\label{sec:introduction}
One of the main challenges of quantitative finance has been to come up with models for stock returns that could reproduce the implied or historical distributions of asset prices, to both acquire knowledge on the underlying dynamics of price formation, and to consistently price and hedge derivative products. Refs.~\cite{Mantegna2000,Bouchaud2003,McCauley2004,Anirban2010} are physicist-friendly references discussing the basics of such problems. 
Seen on a grosser level, finance shares many common features with the study of (unfortunately, not so well-defined) ``complex'' systems, such as the ``random'' nature of the phenomena and the absence of comprehensive and exhaustive theories. The analysis of extreme events plays a pivotal role every time an addressed problem has a stochastic nature, since the rare extreme events can have rather strong or drastic consequences-- making it widely useful in geology, meteorology, as well as in financial economics \cite{Albeverio2006}. Another motivation for studying extreme events in finance can be to account for the observed fat tails of log-returns (deviation from the Normal distribution in the tails) of stock prices. Though the field of extreme statistics is very well-established as part of classical probability theory, its applications can be hindered easily by non-stationarity and, often, slow convergence to the expected results. It is well-known that non-stationarity is the most prevalent cause of anomalous behaviours in financial time series (see for example the discussion in Ref.~\cite{McCauley2008}), and thus the application of extreme event statistics has to be used with caution or reservations in most cases of financial data.

Focusing the analysis on a simple version of an existing financial model, we present how the recently defined simple concept of near-extreme distribution \cite{Sabhapandit2007} can be helpful when studying financial time series data. The poor performance (slow convergence) of the extreme values theory is not totally overcome in this complementary approach. The main aim of this paper is to qualitatively present the possible achievements of this theoretical method. Therefore we use simple standard statistical analyses (Kolmogorov-Smirnoff test and Q-Q plot) to substantiate our results, rather than sophisticated analyses. The paper is organized as follows: Sec.~2 gives a brief review of the main results of the classical theory of extreme values, along with the discussion of the limitations of its application. It also discusses the concept of near-extreme distribution with a related formula. Sec.~3 gives a description of the financial datasets used, together with the explanation of how we model intraday stock returns \cite{Gerig2009} for the demonstration of this approach. Finally, Sec.~4 is dedicated to the results, analyses, discussions and conclusion.

\section{The classical extreme values statistics (EVS) theory}\label{sec:classical}
This theoretical field was born between the 1920s and '30s with the seminal works of von Mises \cite{vonMises1923}, Fr\'echet \cite{Frechet1927} and Fisher \& Tippett \cite{Ficher1928}, and soon became a well-established part of classical statistics with the works of Gnedenko \cite{Gnedenko1943} and Gumbel \cite{Gumbel1958}. Refs.~\cite{Embrechts1997,Kotz2000} are modern and comprehensive introductions of the subject.

The cumulative distribution function $F$ 
of the maximum $x_M$ of a finite set of $N$ $iid$ values $(x_1,x_2,\ldots,x_N)$ distributed respecting the probability density function $g$, may be written as
\begin{equation}
F(x_M) = G(x_M)^N
\label{eq:cdffinite}
\end{equation}
where $G(x)=\int_{-\infty}^x g(u)du$ is the cumulative distribution function of any $x$; sometimes $g$ and $G$ are called \emph{parent distributions}. Note that
here we will continually use only the term ``maximum'' bearing in mind the fact that it actually refers to both the possible extremes, since each observation regarding maximal values is equally true for minimal values. 

The theory states that, in the limit $N\to \infty$, the function $F(x_M)$ converges to either a Weibull, a Fr\'echet or a Gumbel distribution,
\begin{equation}
F(a_N x+b_N) = G(a_N x+b_N)^N \to L(x),\,\,\,\, N\to \infty,
\end{equation}
for some suitable couple of weights $a_N$ and $b_N$ (we will omit the index $M$ from $x_M$ when the context is clear). What discriminates between the three cases of the limiting distribution $L(x)$ is the behaviour of $g$ when $x$ tends to infinity, i.e. the tails. We do not discuss the details of the kind of convergence the theory predicts in different cases; the interested reader can assume the ``conservative'' choice of \textit{weak convergence} (convergence in distribution). Moreover, we do not present an exhaustive description of the domain of attractions, but rather focus on the important cases with a language suited for practical applications.

\subsubsection*{Weibull (Bounded function)}
If the positive support of $g$ is bounded, then $F$ converges to a Weibull distribution $L_W(x)$:
\begin{equation}
L_W(x) = \exp(-(-x)^\beta),\,\,\,\,\, \mbox{with }\beta \ge 1 .
\end{equation}
The appropriate weights $a_N$ and $b_N$ are
\begin{equation}
a_N = b_N - \inf\left\{x: 1-G(x)\le \frac{1}{N} \right\}\,\,\,\mbox{and}\,\,\,b_N = \sup \left\{x: G(x)< 1\right\}
\end{equation}
and the parameter $\beta$ is given by the behaviour of $G$ when approaching the value $w=G^{-1}(1)$: $G(x) \sim (w-x)^\beta$ for $x\to w$. If $w$ is reached with an exponent smaller than unity, then the case is degenerate with a limiting distribution $L_d(x) = \delta(x-w)$.

\subsubsection*{Fr\'echet (Power-law behaviour)}
When $g$ presents a power law behaviour $G(x)\sim x^{-\alpha}$ (for $x\to \infty$ and for $\alpha >0$), the extremal-value distribution approaches the Fr\'echet distribution $L_F(x)$:
\begin{equation}
L_F(x) = \exp(-x^{-\alpha}).
\end{equation}
In this case the weights are given by 
\begin{equation}
a_N = \inf \left\{ x: 1-G(x)\le \frac{1}{N} \right\} \,\,\,\mbox{and}\,\,\,b_N = 0.
\end{equation}

\subsubsection*{Gumbel (Unbounded exponential - or faster - behaviour)}
Finally, when the parent function $g$ has an unbounded support and its behaviour at infinity is exponential (or faster) we recover the Gumbel distribution $L_G(x)$:
\begin{equation}
L_G(x) = \exp(-\exp{(-x)}).
\end{equation}
The weights are given by 
\begin{equation}
a_N = \inf \left\{ x:1-G(x)\le \frac{1}{N e} \right\} - b_N \,\,\mbox{and}\,\,\,b_N = \inf \left\{ x:1-G(x)\le\frac{1}{N} \right\}.
\end{equation}

\subsection{The limitations of EVS in straightforward applications}
Unfortunately, there are some serious limitations in the direct application of this theoretical structure. When dealing with financial time series we inherently work with non-stationary processes, which in this case can be translated into log-returns being non-identically distributed. In other words, $x_i$ and $x_j$ with $i \ne j$, do not have, in general, the same distribution. Moreover, often the convergence to the theoretical distribution is very slow and the extreme values cannot be assumed to be distributed according to $L$, if not for a very high value of the time series length $N$ \cite{Luceno1994,Gyorgyi2008}. However we note that the hypothesis of independence between the elements of the analysed set can be weakened. For example, the same results still hold in the Gaussian case where the correlation between the $x_i$ and $x_{i+k}$ goes to zero as $k^{-\gamma}$ with $\gamma>1$, when $k\to \infty$ \cite{Berman1964,Pickands1969}.

\begin{figure}
\centering
\includegraphics[height=5.5cm, width=1.05\columnwidth]{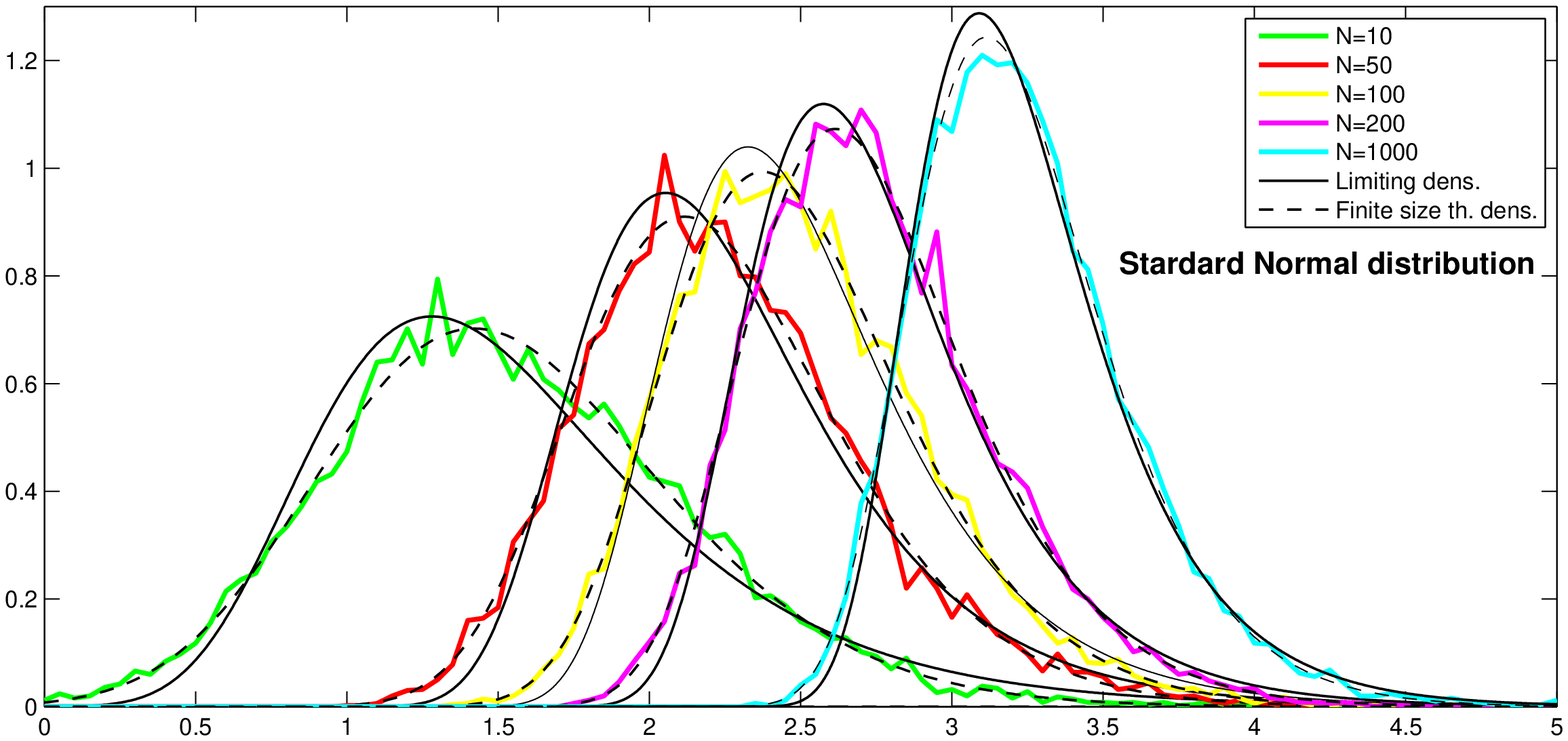}
\includegraphics[height=5.5cm, width=1.05\columnwidth]{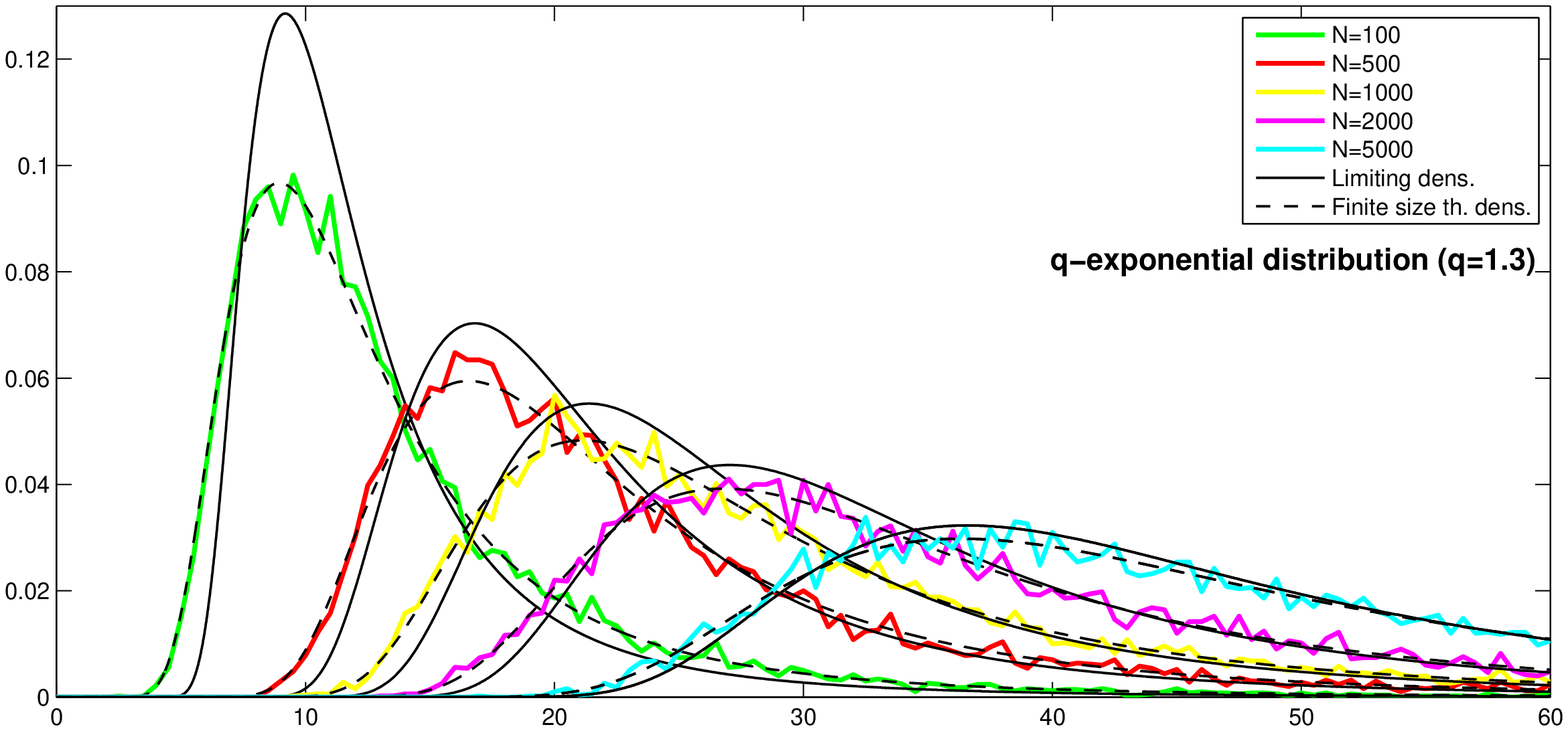}
\caption{Two explicit examples of slow convergence to the limit distribution. Top: standard Normal distribution. Bottom: $q$-exponential distribution with index $q=1.3$ ($\eta = - 1/0.3$, see text after Eq.~\ref{eq:qexponential}). The color lines are the empirical densities of maxima for synthetic variables respecting the two different parent distributions (the sample size of each statistics is 1000). The solid black lines are the limiting densities $L_G((x-b_N)/a_N)/a_N$ and $L_F((x-b_N)/a_N)/a_N$ and, finally, the black dashed lines are the theoretical finite sample densities.}
\label{fig:qgauss}
\end{figure}

For the illustration of the slow convergence, we use two important distributions: Gaussian and Tsallis $q$-exponential. The Gaussian (standard Normal) distribution is ubiquitous, and its definition or properties well-known, and hence the choice. The choice of the latter, because of many recent applications \cite{Tsallis2003,Borland2004,Politi2008} and its gaining usefulness. For the Tsallis $q$-exponential distribution, we have 
\begin{equation}
g(x) =  (1+(q-1)x)^\frac{q}{1-q}\,\,\,\mbox{for}\,\,\,x>0\,\,\mbox{with}\,\,q>1;
\label{eq:qexponential}
\end{equation}
it can be alternatively defined for a larger range of the parameter $q$, but we focus on this particular case. In econometrics, the same distribution is known as the generalized Pareto law, and its tail index is equal to $\eta=1/(q-1)$. Thus, its limiting distribution is the Fr\'echet one.
Fig.~\ref{fig:qgauss} depicts the extreme distribution for both Gaussian and $q$-exponential variables, obtained using simple pseudo random generation. From the figure, it is clearly evident that even for $N$ in the order of thousands, the difference between finite size theoretical distributions and the theoretical limiting distributions is still large, and that the results of synthetic data are in fair agreement with the finite size distribution, as ought to be. The mismatch could arise from the weights $a_N$ and $b_N$; it is possible that the shape is already satisfactorily close to the limiting one, but this effect is not clear because of a slow convergence in the weights. However, there is no consistent way to discriminate the nature of the slow convergence, and to our knowledge no ``finite size adjustments'' for the weights are present in the literature.

\begin{figure}
\centering
\vspace{-1cm}\includegraphics[ width=1.05\columnwidth]{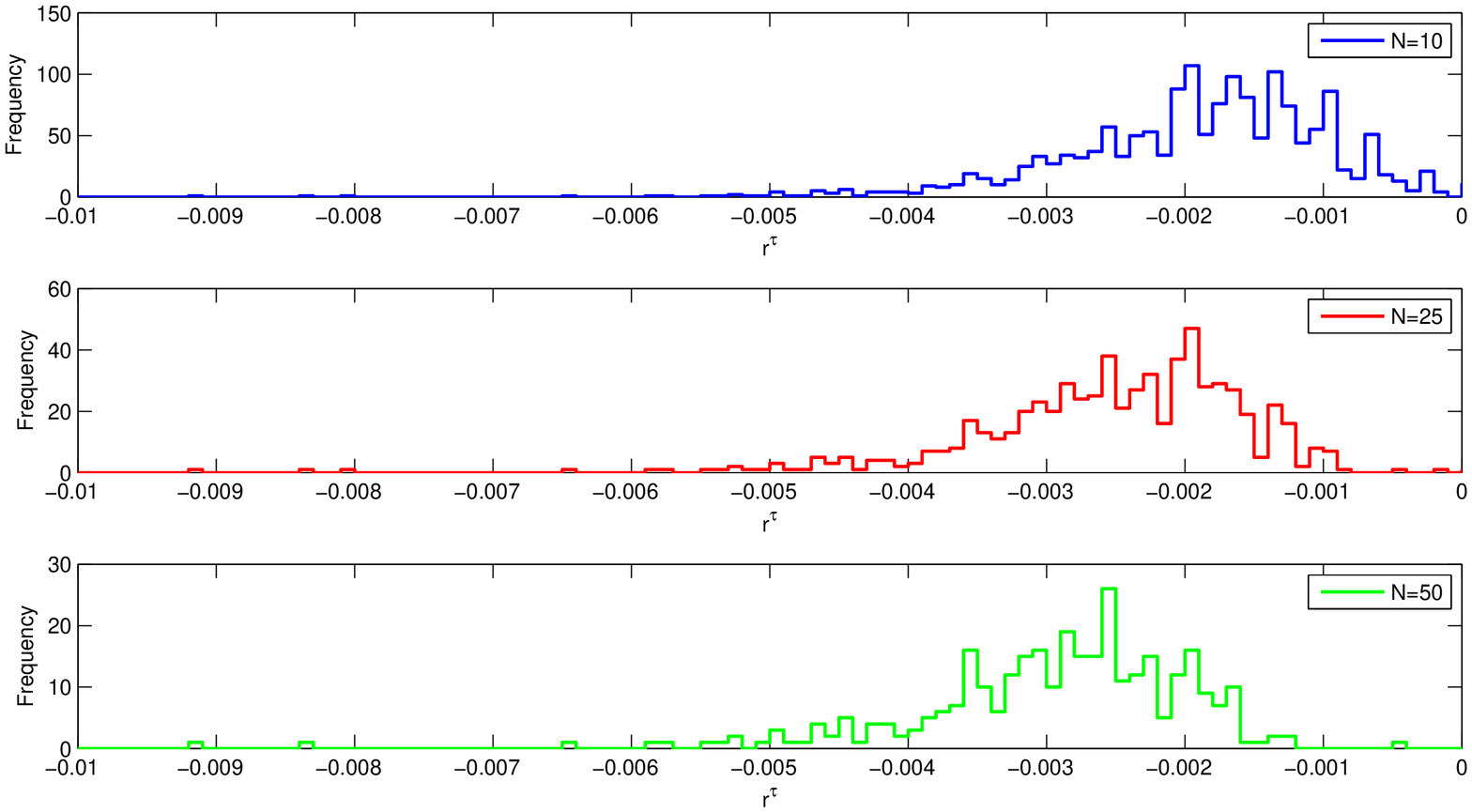}
\caption{Minimal values distributions for the MSFT data for $\tau=2500$ and different values of $N=10,25,50$, highlighting the effect brought by discretization. For the full explanation of the symbols, legend and parameters $\tau$, $N$ please refer to Sec.~\ref{sec:qualcosa}. The bin size is equal to $10^{-4}$.}
\label{fig:minima}
\end{figure}

Finally, we must also add the pragmatic issue that arises from the discrete nature of prices in the market. \textit{Per se}, this visible mis-match does not undermine the application of the theory, but in order to make the function ``smooth'' and observe a satisfactory convergence, a very large value $N$ would be required again, enormous for the case in consideration. Fig.~\ref{fig:minima} shows the raw statistics of minimal values for market data, illustrating the issue.

\subsection{The near-extreme distribution}
The idea that helps to overcome the mentioned problems is to analyze a closely related distribution, the near-extreme distribution, recently proposed in Ref.~\cite{Sabhapandit2007}, and already applied to similar problems in Ref.~\cite{Lin2010}. Roughly speaking, the near-extreme distribution is the distribution of the distance from the maximal value in a finite set. Considering again a finite set of $N$ $iid$ values $(x_1,x_2,\ldots,x_N)$ distributed according to the parent density $g$ (or its cumulative $G$), we call their maximum $x_M=\mathrm{max}(x_1,x_2,\ldots,x_N)$. According to Ref.~\cite{Sabhapandit2007}, the empirical near-extreme density with respect to the maximum is defined as

\begin{equation}
\rho_e(r,N) = \frac{1}{N-1} \sum_{x_i \neq x_M} \delta \left[ r - \left( x_M - x_i\right) \right],
\label{eq:dos}
\end{equation}
where $r$ is the distance measured from the maximum value $x_M$, and $x_M$ is not counted itself. Note that in order to obtain $\int_0^\infty \rho_e(r,N)dr = 1$ we are using a different normalization than in Ref.~\cite{Sabhapandit2007}.

Under the assumptions and with the current notations we have introduced above, we can obtain the following expression for the expected density
\begin{equation}
\rho(r,N) = \int_{-\infty}^{+\infty} N g(x) G(x)^{N-2} g(r-x) dx.
\end{equation}
This can be justified noticing that $\rho(r,N)$ is the convolution of the density of the distance from a given maximum with the pdf of the maximum; the former can be written as $g(x-x_M)/G(x_M)$ and the latter as $Ng(x_M)G^{N-1}(x_M)$ (the derivative of Eq.~\ref{eq:cdffinite}).


In Ref.~\cite{Sabhapandit2007}, the authors describe the property of $\rho(r,N)$ when $N$ goes to infinity and show that near-extreme density converges to different limiting forms depending on the tail of the original distribution. However, we work here with finite sample $N$ only and do not take interest in the limiting forms.

\section{The data and model}
\label{sec:qualcosa}
\subsection{Data set}
\label{sec:data}

The original data set consists of all trades registered in the primary
markets of the analyzed stocks. The data are stored in the Thomson Reuters RDTH data base made available to the Chair of Quantitative Finance by BNP Paribas.
For the purpose of our study, we extract from the RDTH database records consisting in the time of a transaction,
the bid and ask prices prior to each transaction, and the
traded price. These data, appropriately filtered in order to remove
misprints in prices and times of execution, correspond to the trades
registered at NYSE or at NASDAQ during 2007, for four shares of
the Dow Jones Industrial Average Index at that time, namely: C.N, GE.N, INTC.O and MSFT.O. The C.N and GE.N were primarily traded at NYSE, while INTC.O and MSFT.O were primarily traded at NASDAQ. The full meaning of the symbols is available from {\tt www.reuters.com}.
The choice of one year of data is a trade-off between the necessity of managing enough data for significant statistical analyses and the goal of minimizing the effect of strong macro-economic fluctuations. However, the consistency of the discussed results during extreme condition periods are beyond the purposes of the present paper, and are left for future studies. The stocks were chosen among the most active at that time, since, as already discussed, the size of the statistics can pose significant problems. We do not present results on a large set of different stocks because our main purpose is to show the approach rather than the final ``fit''. An extensive study would need a formal definition of approximation rather than a qualitative and general one. Moreover, we are aware the method has to be finely tuned on the single stock characteristics (activity, price, particular trends, etc.) and the raw or straightforward application of it could fail in general. For this reason we show the result on the simplest situations possible: Four similarly liquid stocks in the same index and in the same economy (but not in the same sectors: Finance for C.N, general industry for GE.N and advanced technology for INTC.O MSFT.O).

For each day the considered period is $10:00-15:45$ hrs, precisely. The choice of the considered periods is to restrict the hours only to the central part of the trading day, discarding the opening and closing period. This is justified, since data often exhibit less ``anomalies'' during these parts of the trading day-- errors tend to occur more often during the first and last part of the continuous trading day (it often happens that some shares are opened for trading several minutes after the others, due to potential issues during the opening auction). 

Note that we do not use ``physical time'' as our unit of measure, but rather consider ``trading time'' (a.k.a. ``event time'' or ``tick time'') \cite{Chakraborti2011}. It is incremented each time an ``event'' occurs, that is each time the mid-price changes, i.e., each time \textit{either} the bid \textit{or} the ask price changes. In this way, we do not consider a trade leaving the mid-price unchanged as an event. As described in the following subsection, the final analysis is then performed by aggregating (summing) these events. 

\subsection{The intraday model and approximation idea}
\label{sec:idea}
Given $S_i$, the price time series sampled in ``event time'', and a time-lag $\tau$, the log-returns $r^\tau_i=\ln \frac{S_i}{S_{i-\tau}}$ may be modeled \cite{Gerig2009} as a discrete-time stochastic process with a fluctuating variance:
\begin{equation}
r^\tau_i = \sigma_i \epsilon_i,
\label{eq:idea}
\end{equation}
where $\sigma_i^2$ is the local variance of the process, and $\epsilon_i$ are samples of a standard Normal distribution. 
Any drift for the returns may be neglected for the time scales under consideration.  
It is assumed that $\sigma_i$ is varying slowly enough, so that it can be treated as a \textit{constant} over intraday time scales. 
Replacing $\sigma_i$ with its local constant value $\sigma$, individual returns can be approximated as $r^\tau_i\approx \sigma \epsilon_i$. Eq.~\ref{eq:idea} is a simplified version of ARCH-GARCH-like models used by the authors of Ref.~\cite{Gerig2009} to fit market fluctuations using simple statistics of the squared returns. As often happen in econometrics this leads to a \textit{mixture of Normal distributions}, helping us to model the non-stationarity and to overcome the tedious problem of evaluating the tail index of a candidate parent distribution $G$ \cite{Goldstein2004,CoronelBrizio2010}. For the sake of clarity, we want to stress the fact we call ``time-lag'' or ``time-window'' the parameter $\tau$ but it is not immediately connected with the physical time.

In our studies we proceed as follows:
\begin{itemize}
\item Fix a positive integer $\tau$, representing a time-lag.
\item From the original price time series $S_t$ we extract the $r^\tau_i$s defined as 
\begin{equation}
r^\tau_i = \log \frac{S_{(i+1)\times\tau}}{ S_{i\times\tau}}\nonumber
\end{equation}
without allowing any overlap.
\item Arrange consecutive $r^\tau_i$s in sets of length $N$, labeled with the index $j$ that runs from $1$ to $h$, such that $h$ is the number of sets obtained and the total time series length $T=h \times N$. Within each set $j$, we estimate $\sigma_j^2$ as the classical sample variance, and also evaluate the near-extreme statistics with respect to its maximum $x_M^j$, according to \eq{eq:dos}. 
\item The empirical near-extreme statistics is then given by the aggregation of each one of those statistics:
\begin{equation}
\rho_e(r,N) =  \frac{1}{h} \sum_{j=1}^h \frac{1}{N-1} \sum_{x_i \neq x_M^j} \delta \left[ r-\left( x_M^j - x_i\right) \right].
\label{eq:edos}
\end{equation}
\end{itemize}

After some straightforward algebra and using our standing assumption we expect the near-extreme distribution to be
\begin{equation}
\rho(r^\tau ,N) = \frac{1}{h}\sum_{j=1}^h \int_{-\infty}^{+\infty} N g_j(x) G_j(x)^{N-2} g_j(r^\tau - x) dx.
\label{eq:approxmargin1}
\end{equation}
where
\begin{equation}
g_j(x)=\mathcal{N}(0,\sigma_j)\,\,\,\mbox{and}\,\,\,G_j(x)=\frac{1}{2}\left(1+\mathrm{erf}\left(\frac{x}{\sqrt{2}\sigma_j}\right)\right).
\end{equation}
The corresponding cumulative distribution $P$ is given by
\begin{equation}
P(r^\tau) = \frac{1}{h}\sum_{j=1}^h \int_{-\infty}^{+\infty} N g_j(x) G_j(x)^{N-2} G_j(r^\tau - x) dx.
\label{eq:approxmargin_cum}
\end{equation}
In words, we get that the near-extreme distribution of log-returns over a time-window $\tau$ can be obtained as a mixture of the corresponding near-extreme distribution for $h$ Gaussian variables, and that each considered set of length $N$ brings a single element in the mixture.

If the model and the underlying assumptions are valid we should observe an agreement between the expected ``theoretical'' distribution in Eq.~\ref{eq:approxmargin1} and the empirical distribution corresponding to Eq.~\ref{eq:edos}.

\section{Results and discussion}

\begin{figure*}
\centering
\includegraphics[height=5.5cm, width=.49\columnwidth]{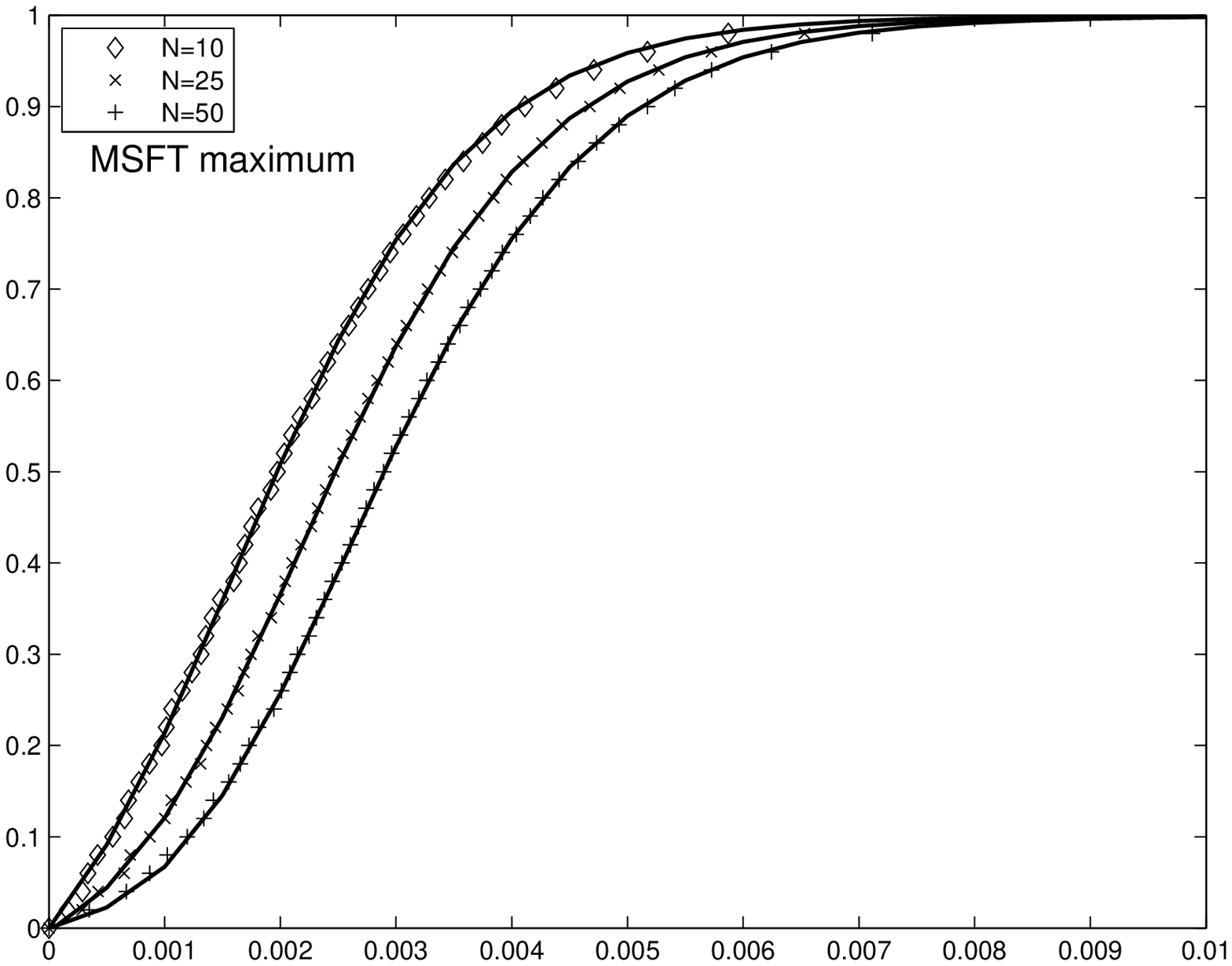}
\includegraphics[height=5.5cm, width=.49\columnwidth]{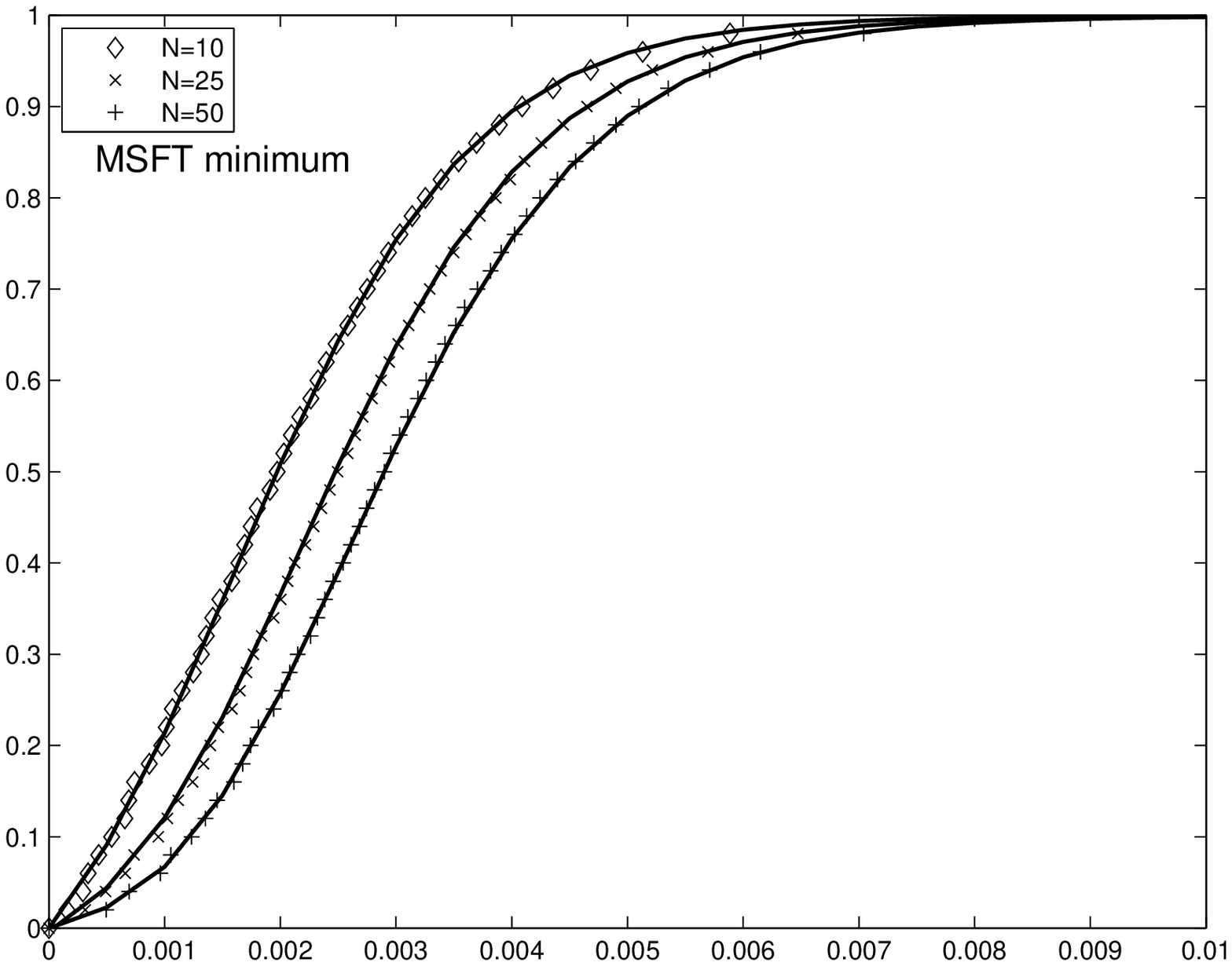}
\includegraphics[height=5.5cm, width=.49\columnwidth]{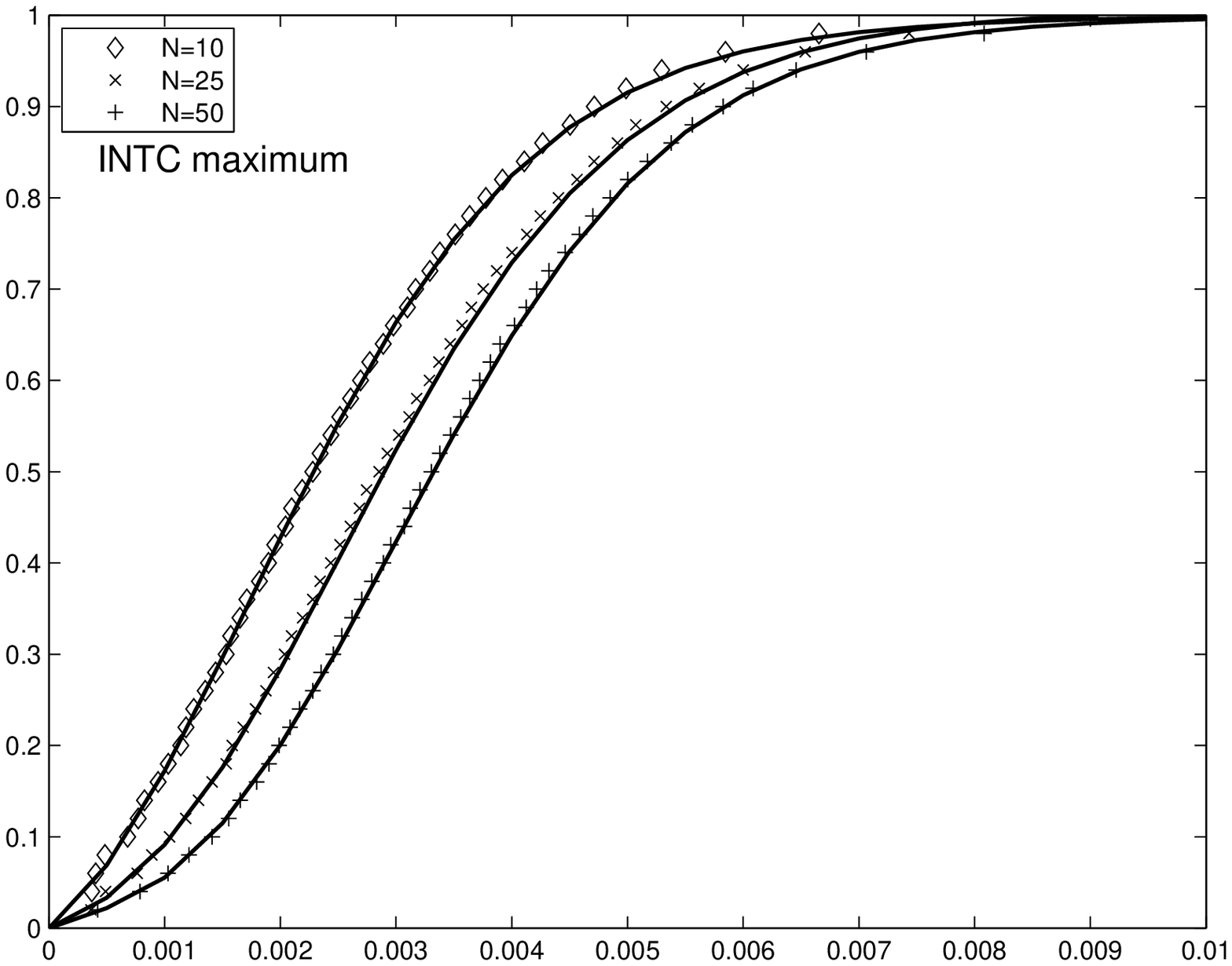}
\includegraphics[height=5.5cm, width=.49\columnwidth]{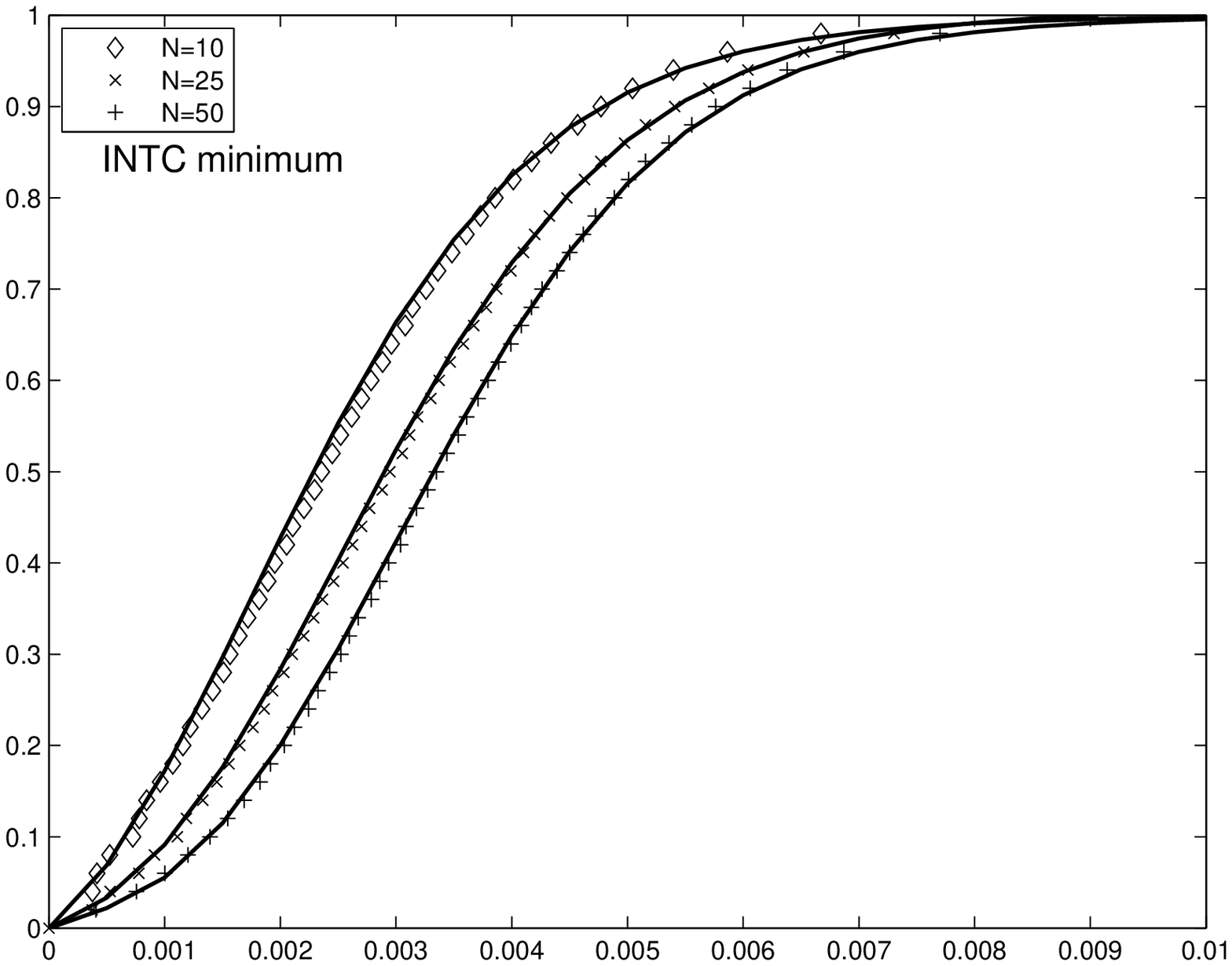}
\includegraphics[height=5.5cm, width=.49\columnwidth]{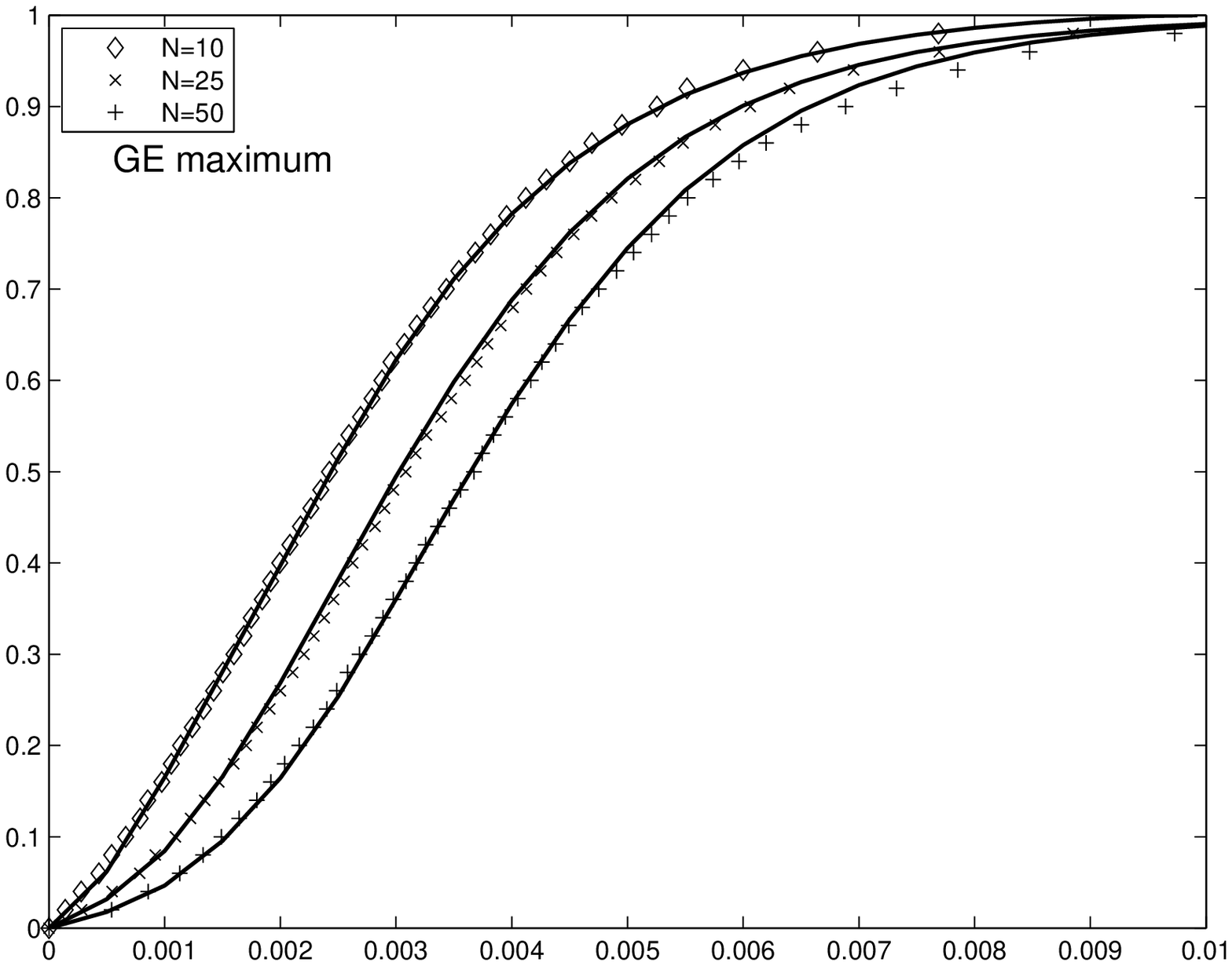}
\includegraphics[height=5.5cm, width=.49\columnwidth]{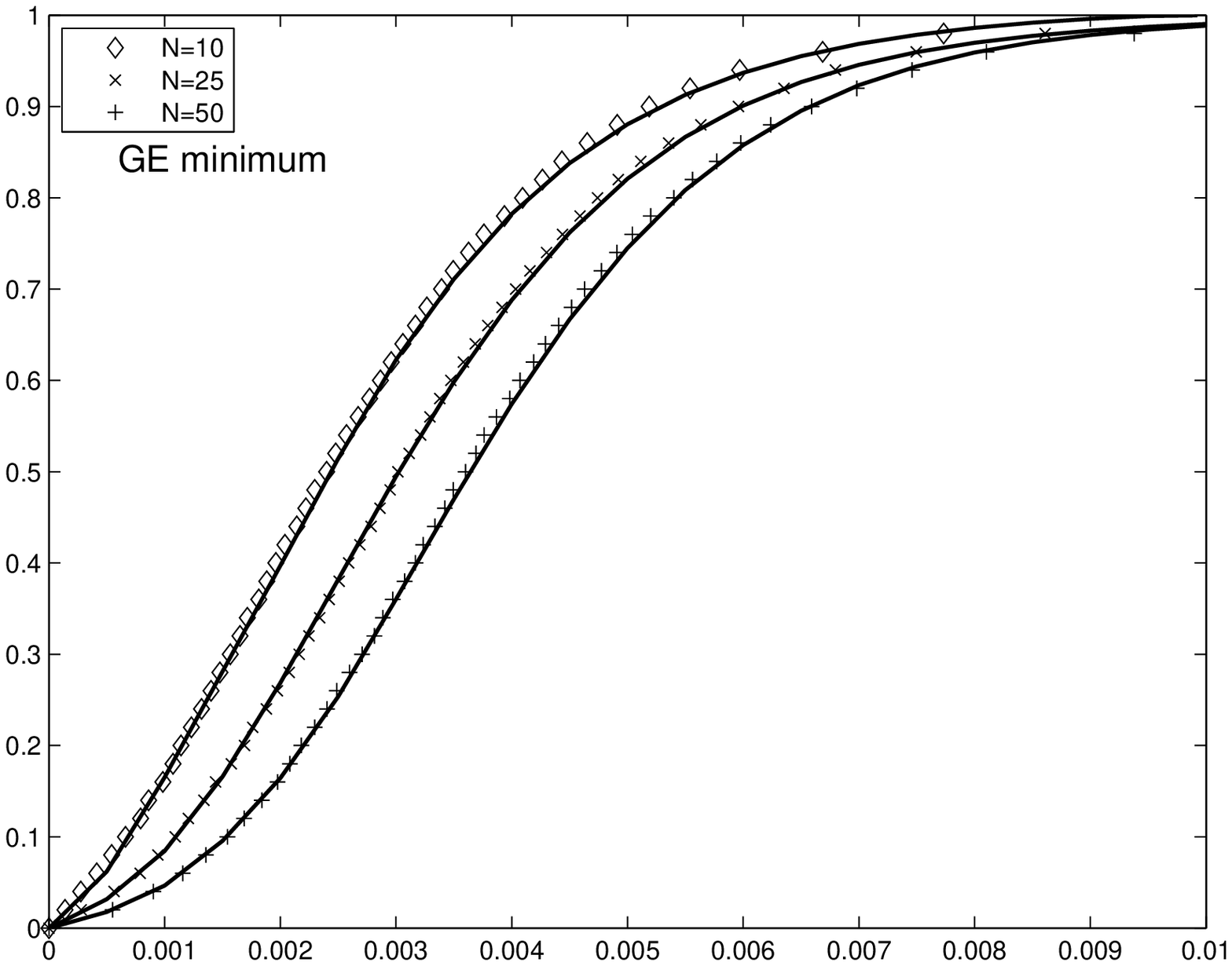}
\includegraphics[height=5.5cm, width=.49\columnwidth]{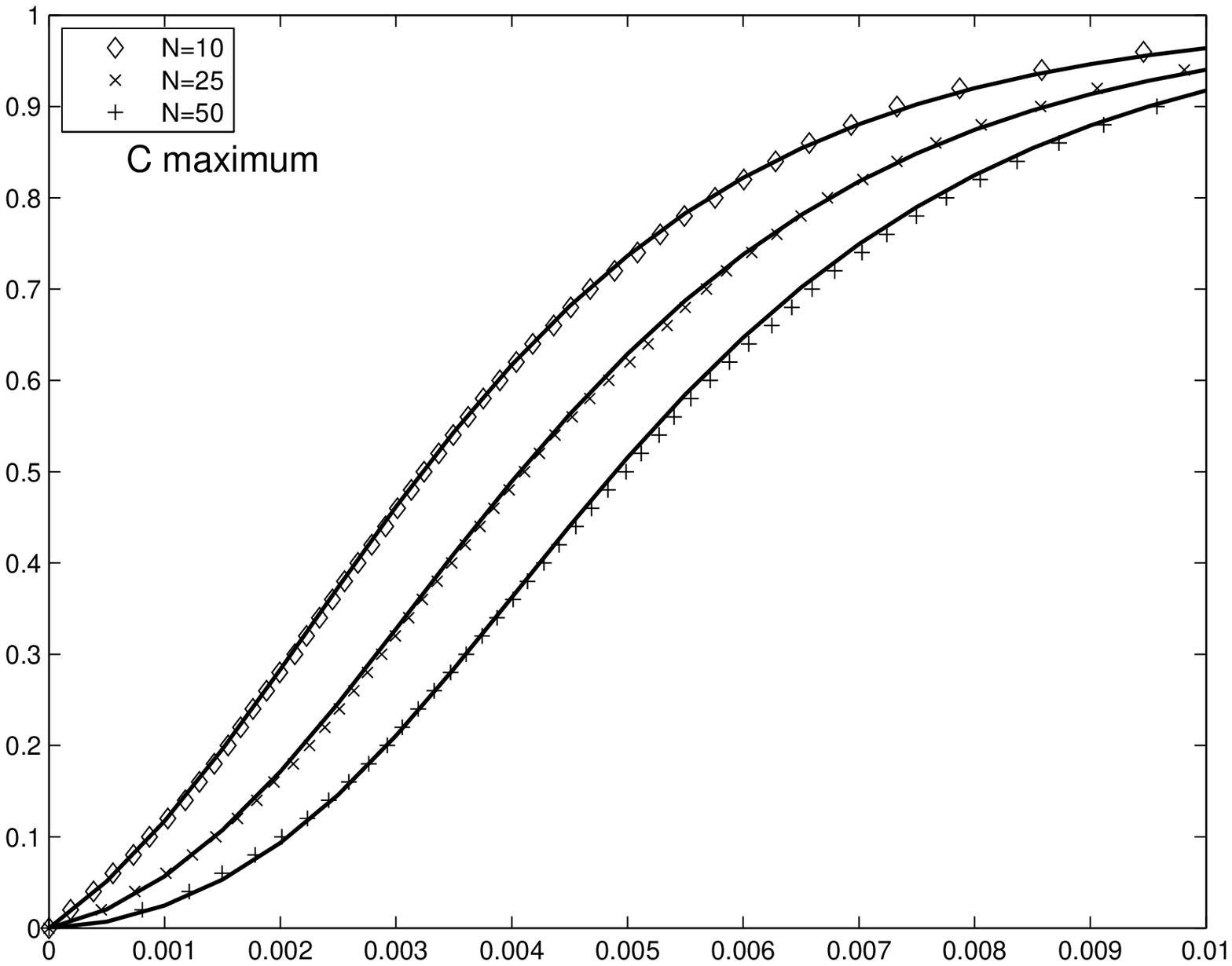}
\includegraphics[height=5.5cm, width=.49\columnwidth]{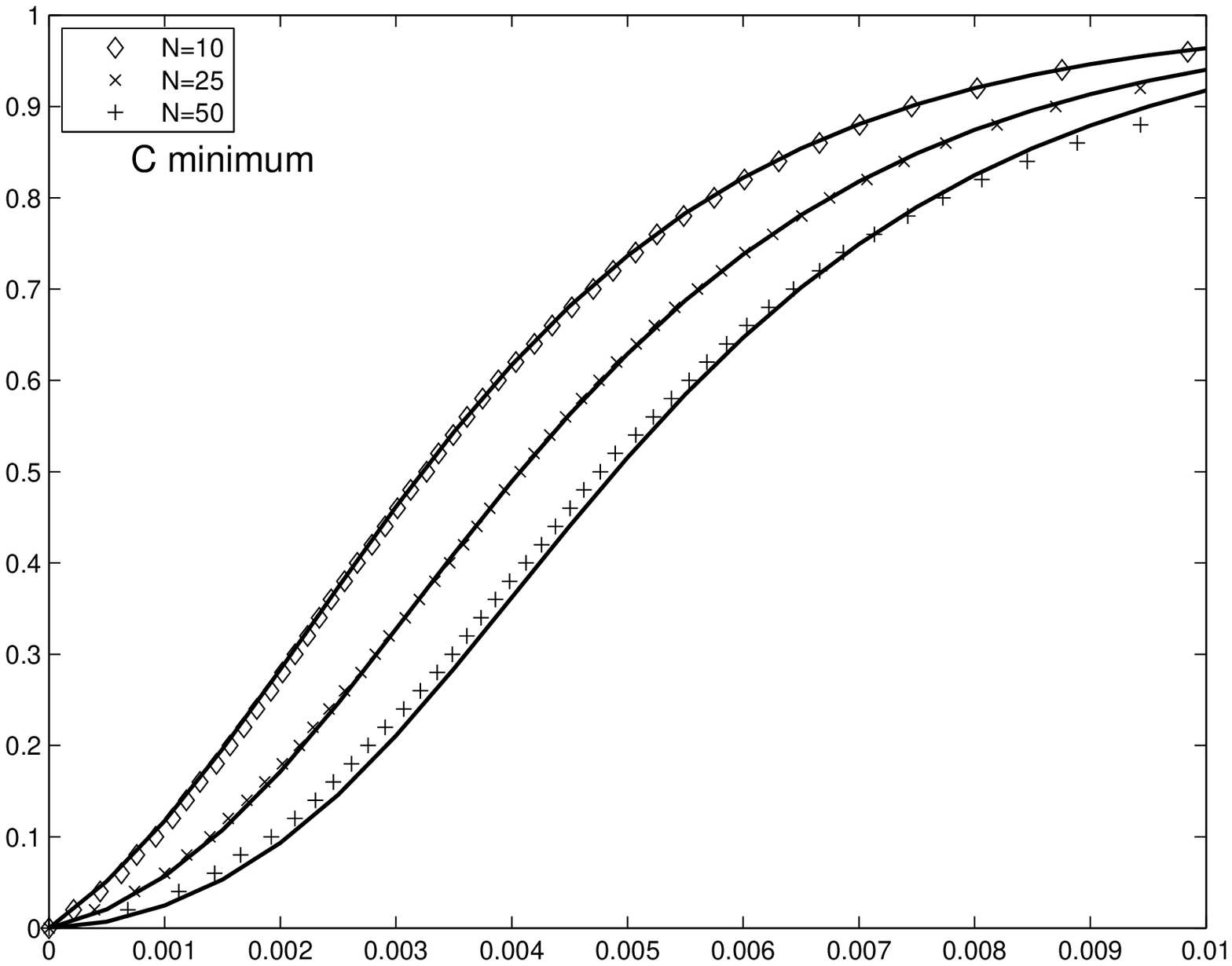}
\caption{Examples of empirical near extreme distribution vs. the expected theoretical distribution for the four analyzed stocks. The considered parameters are $\tau = 2500$ and $N=10,25,50$.}
\label{fig:fitnearex}
\end{figure*}

\begin{figure*}
\centering
\includegraphics[height=5.5cm, width=.49\columnwidth]{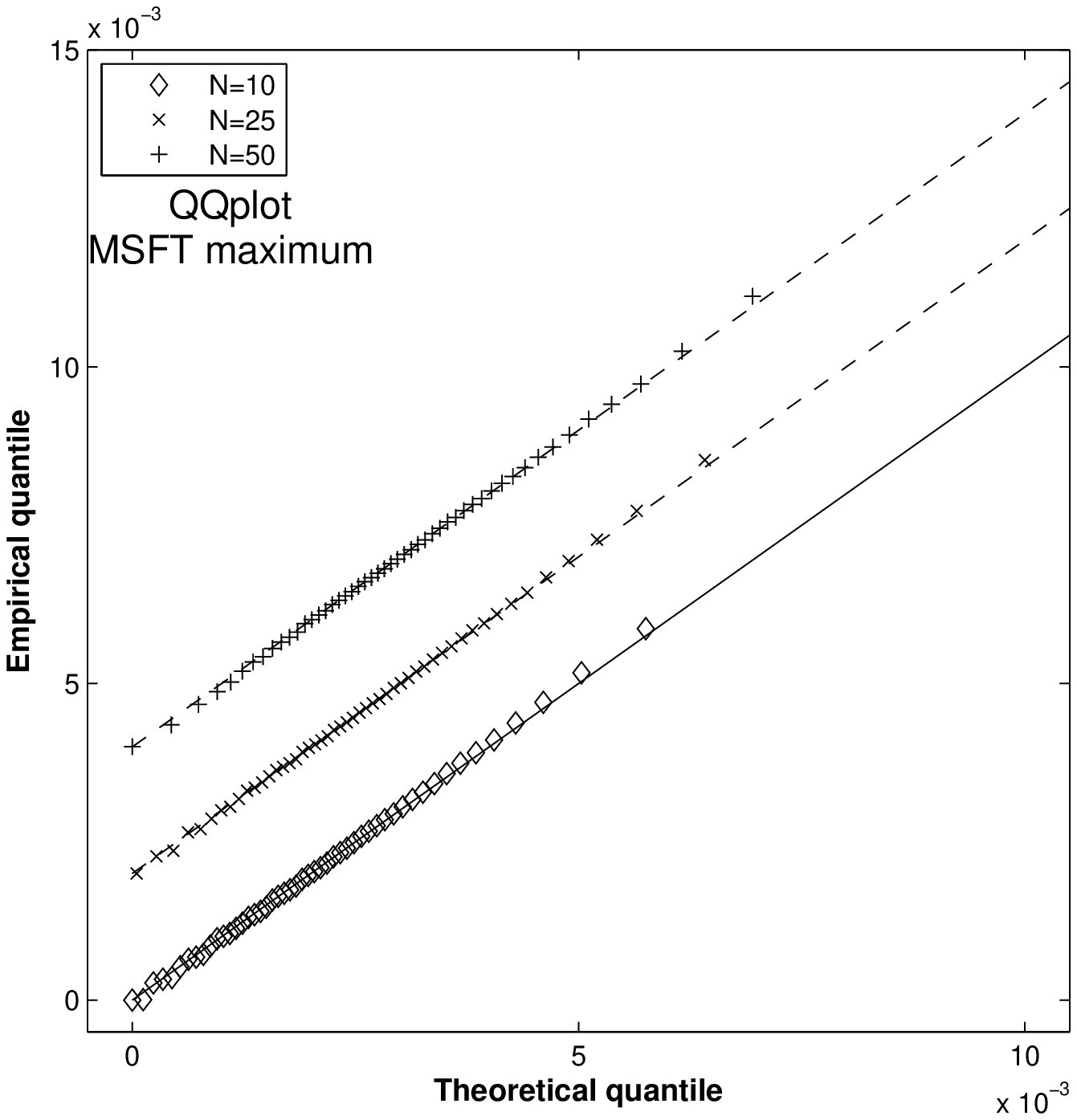}
\includegraphics[height=5.5cm, width=.49\columnwidth]{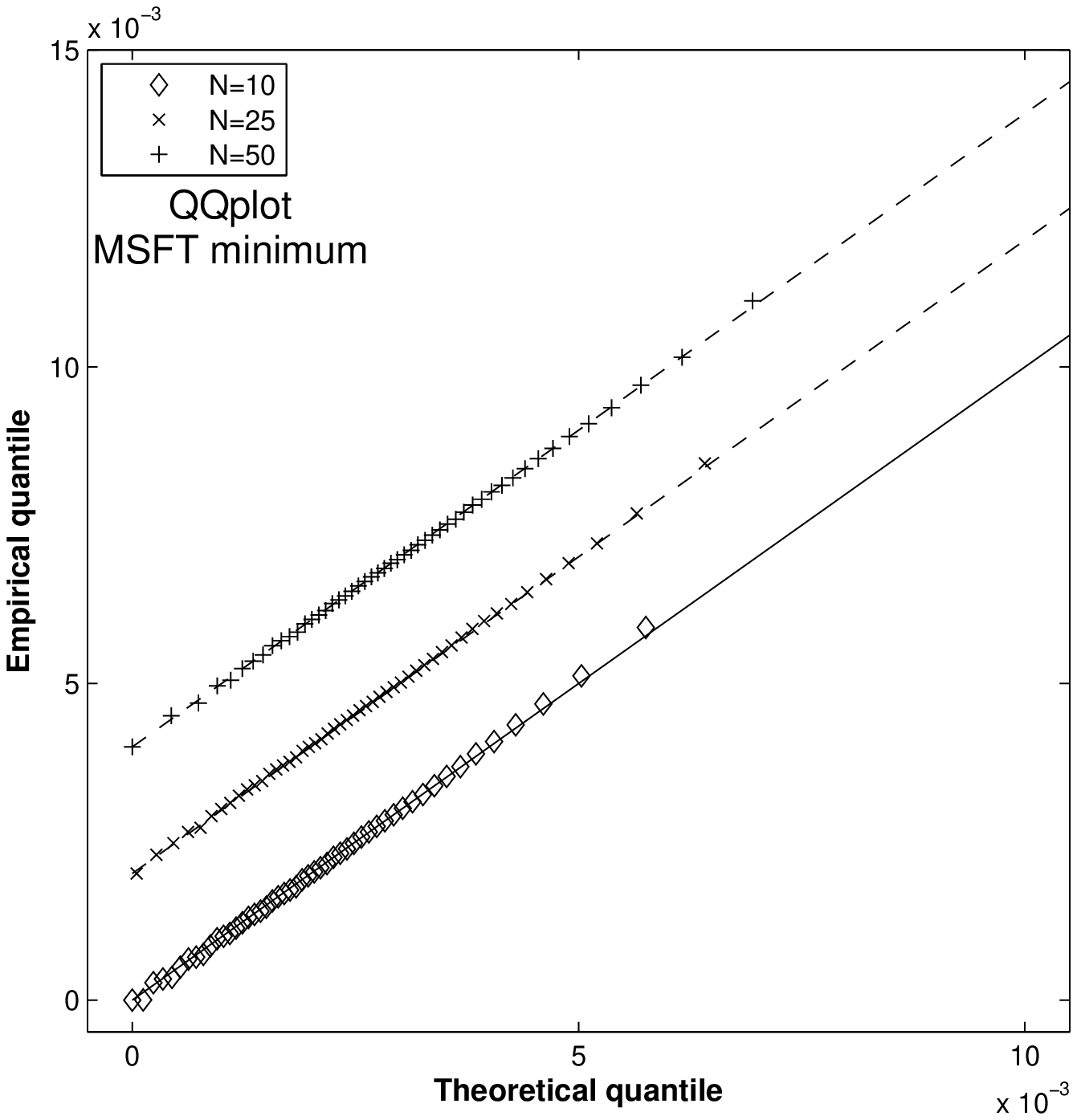}
\includegraphics[height=5.5cm, width=.49\columnwidth]{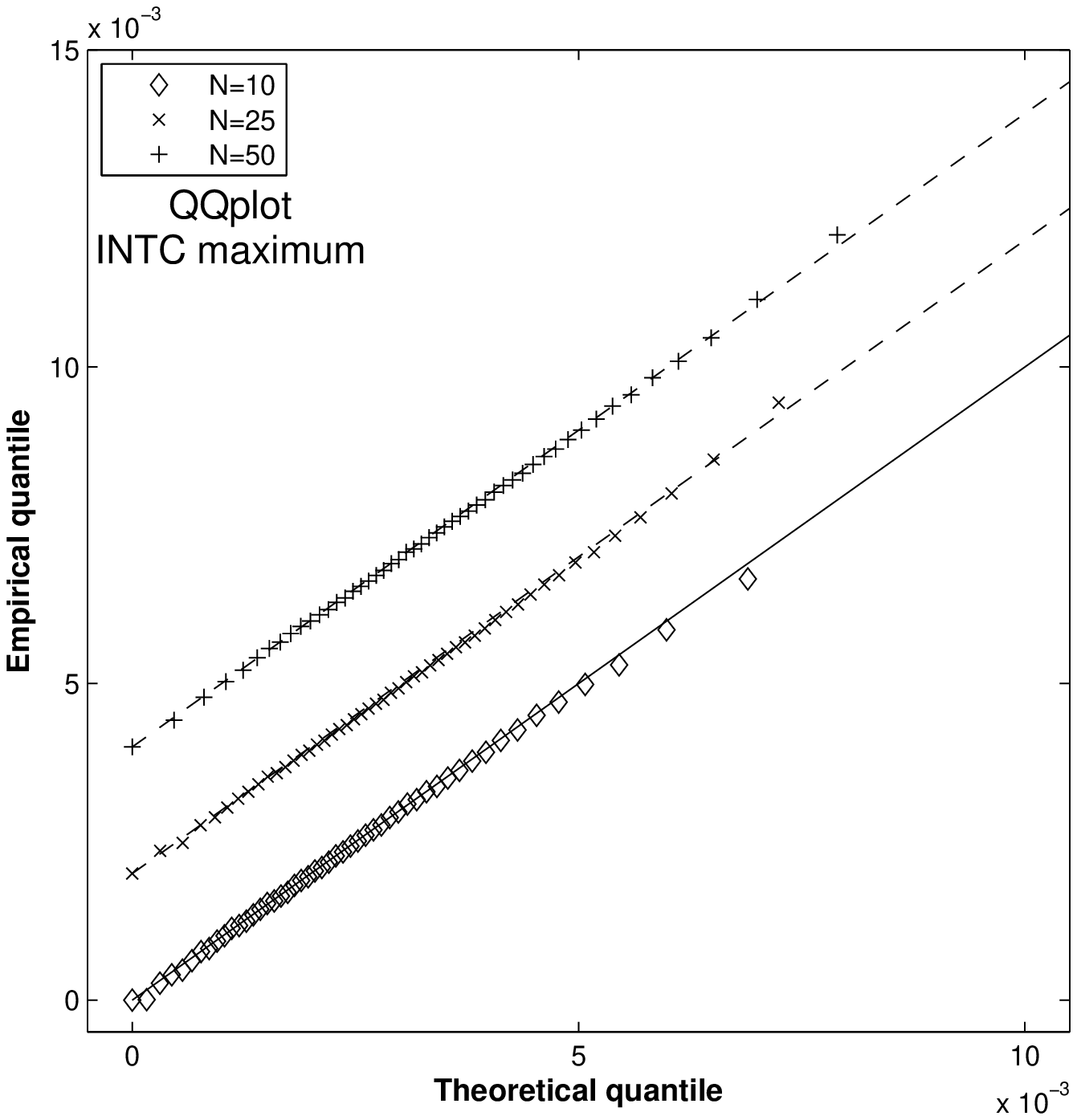}
\includegraphics[height=5.5cm, width=.49\columnwidth]{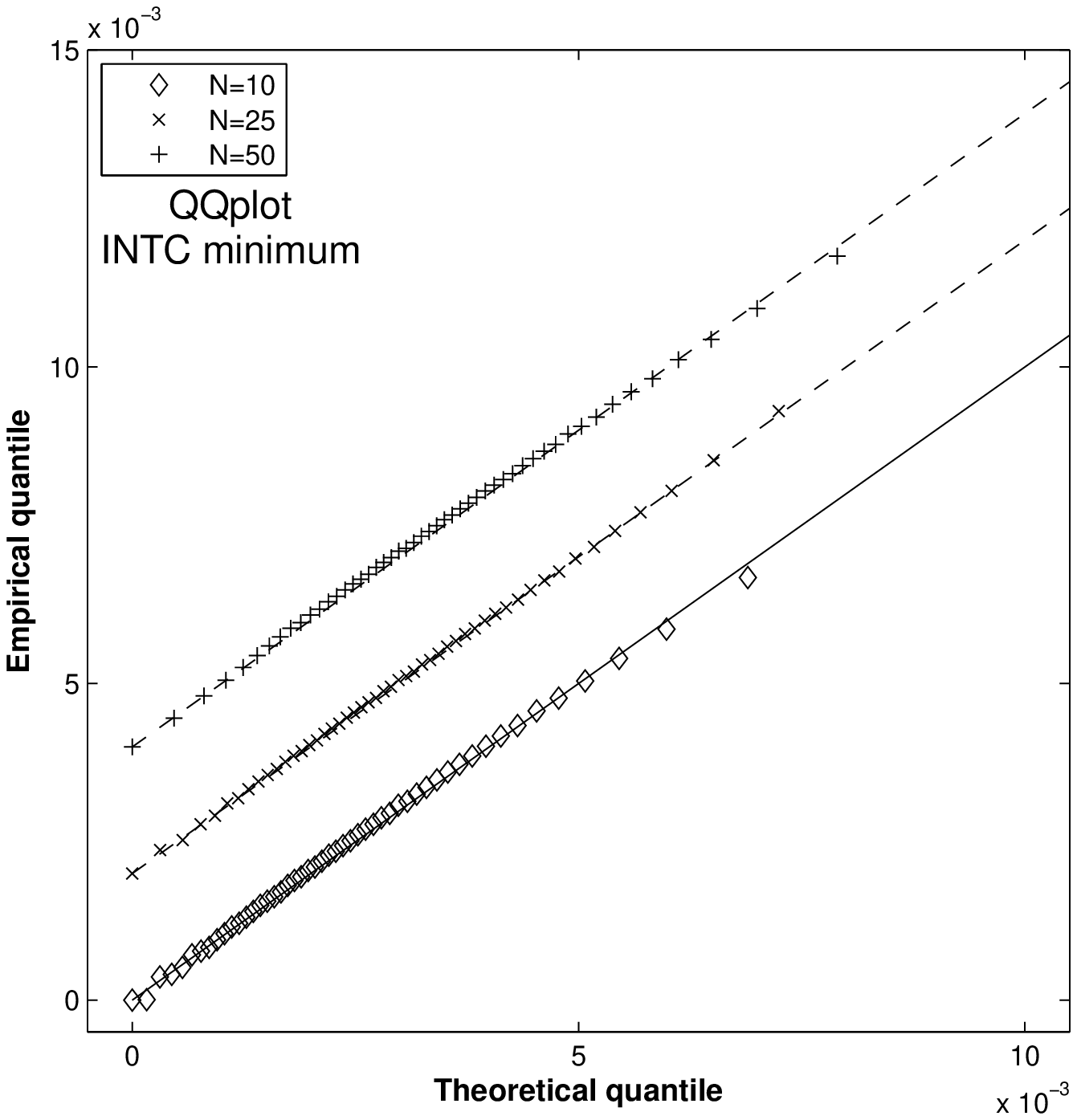}
\includegraphics[height=5.5cm, width=.49\columnwidth]{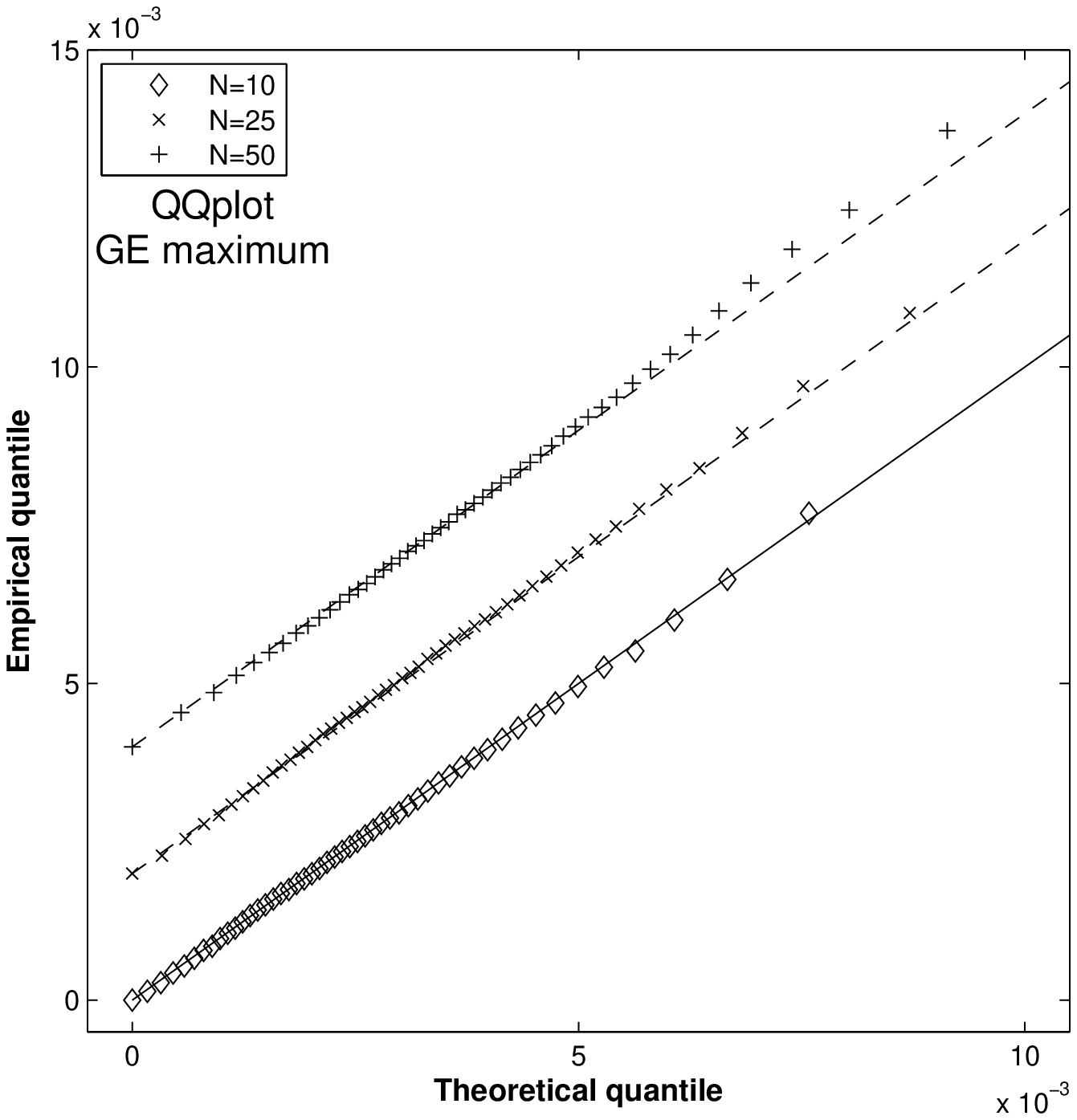}
\includegraphics[height=5.5cm, width=.49\columnwidth]{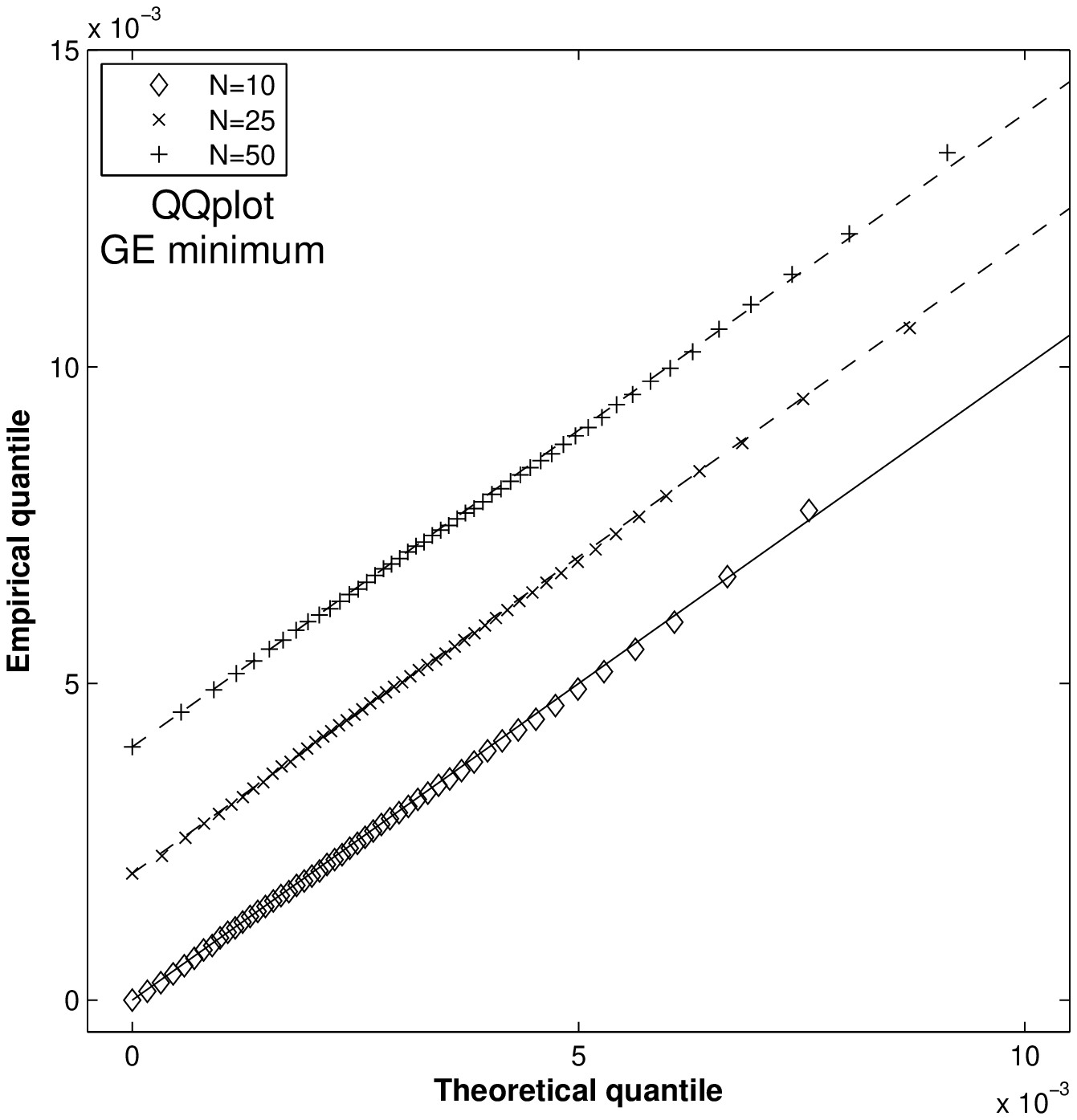}
\includegraphics[height=5.5cm, width=.49\columnwidth]{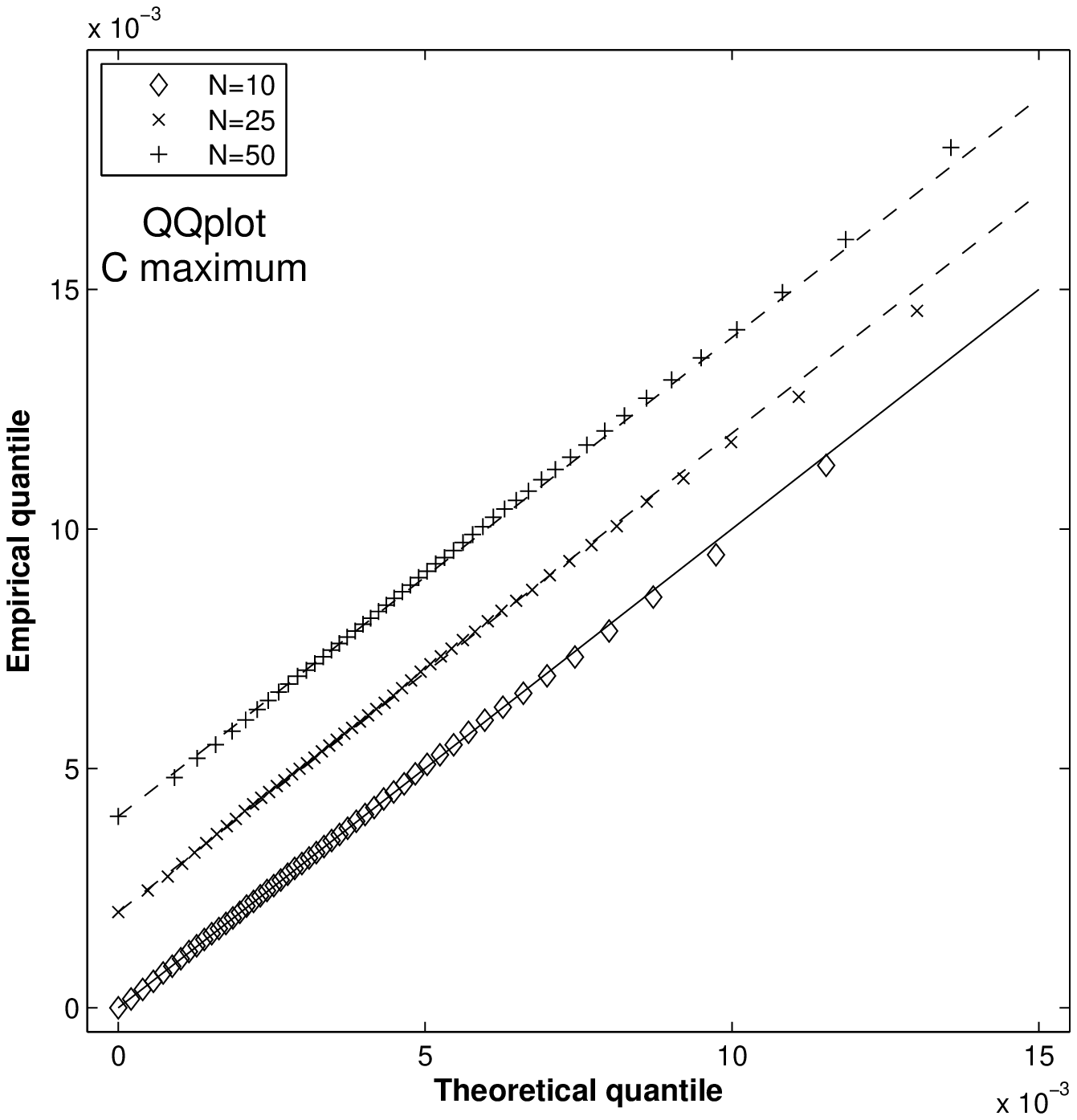}
\includegraphics[height=5.5cm, width=.49\columnwidth]{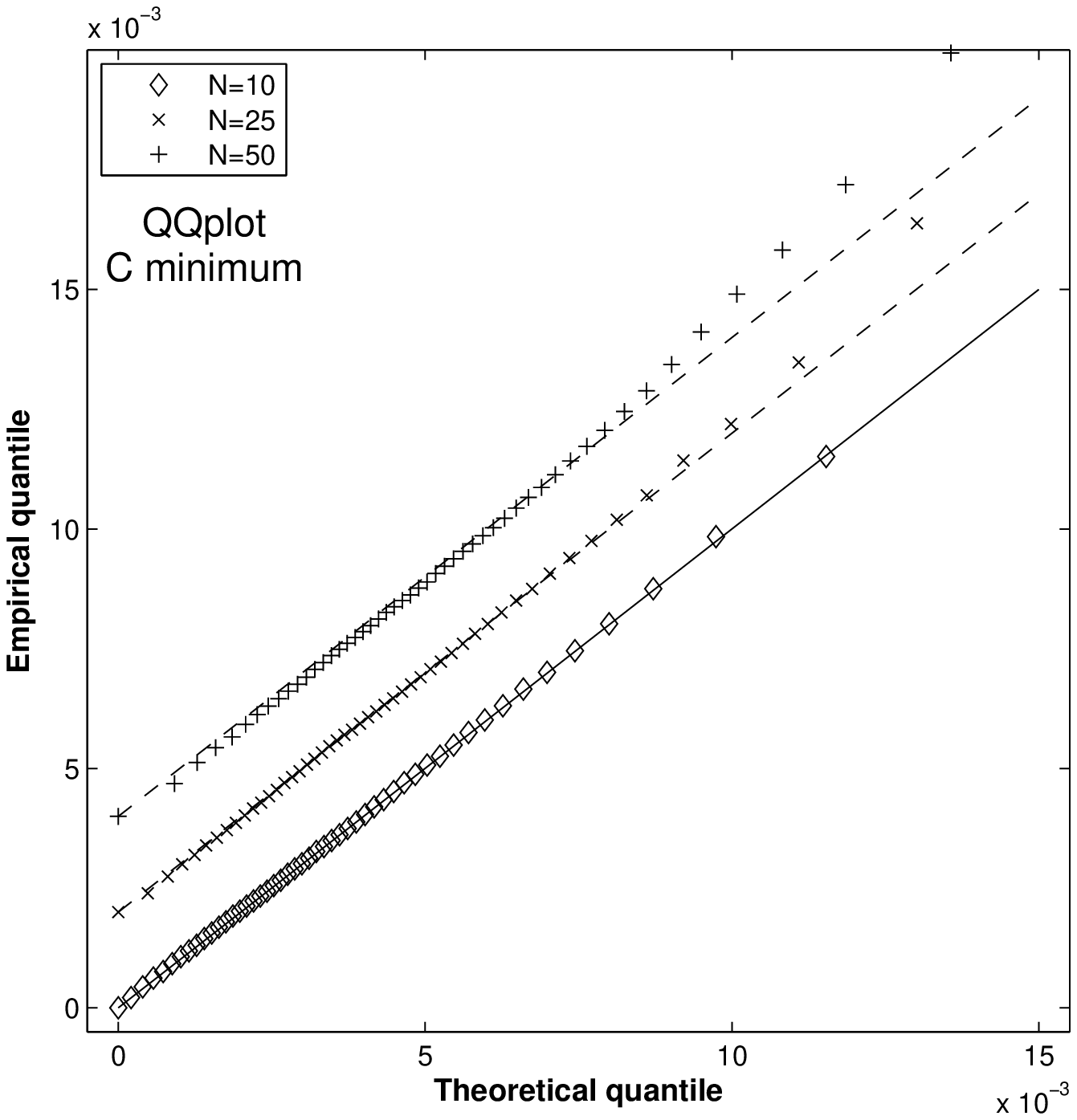}
\caption{Q-Q plots for the near-extreme statistics depicted in Fig.~\ref{fig:fitnearex}. The match between theoretical and empirical quantiles is satisfactory, except for some differences approaching 1 (upper-right part of plots). The cases $N=25$ and $N=50$ have been shifted for reader convenience. The considered quantiles start zero to 0.98 with a step of 0.02.}
\label{fig:QQplot}
\end{figure*}

\begin{table}[thb]
\begin{center}
\begin{tabular}{ccccc}
     & C max & C min & MSFT max  & MSFT min\\
     \hline
N=10 & 0.493 & 0.678 & 0.923     & 0.961 \\
N=25 & 1.06  & 0.784 & 0.602     & 1.137 \\
N=50 & 1.23  & 2.03**& 1.029     & 0.595 \\
 & & & &  \\
     & INTC max & INTC min & GE max & GE min\\
     \hline
N=10 & 1.524*  &  2.320** &  0.560 & 1.228 \\
N=25 & 1.567*  &  1.673** &  1.226 & 0.706 \\
N=50 & 0.954   &  2.697** &  1.520*& 1.022 \\
\end{tabular}
\end{center}
\caption{K-S statistics for the distributions depicted in Fig.~\ref{fig:fitnearex}. One star stands for the fail of the test at the 5\% significance level (critical value 1.333), two stars for the fail at 1\% (critical value 1.625). The values confirm the quality of the fit for all cases but INTC, where we experience some problems, especially when analyzing the near extreme distribution with respect to the minima.}
\label{tab:ks}
\end{table}

\begin{table}[thb]
\begin{center}
\begin{tabular}{ccccc}
     	& C     & GE  &  MSFT       &  INTC\\
     \hline
N=10 	& 7946  & 7206  & 14496     & 14398 \\
N=25 	& 8472  & 7679  & 15501     & 15380 \\
N=50 	& 8624  & 7840  & 15826     & 15677 \\
\end{tabular}
\end{center}
\caption{Sample sizes; the K-S statistics are obtained from the formula $\sqrt{\mbox{Sample\,\,size}}\max\|P(r^\tau) - P_e(r^\tau)\|$.  }
\label{tab:sample}
\end{table}

In 2007, NYSE and NASDAQ had 251 days of open business and in the analyzed periods (see Sec.~\ref{sec:data}) we isolated more than 22.4 [million] events for C, 20.3 for GE, 40.4 for INTC and 40.8 for MSFT. Fixing $\tau = 2500$ we obtain more than 30 intraday returns for C and GE and around 65 for INTC and MSFT. With these parameters a set of $N$ returns could be a bit longer than a day, but its components remain inherently intraday objects. Fig.~\ref{fig:fitnearex} shows the main results obtained considering $N=10,25,50$. The solid lines are the ``theoretical'' distributions as defined by the discussion in the previous sections (Eq.~\ref{eq:approxmargin1}) and the dots report the direct estimation of the near-extreme cdf.

The agreement between the theoretical and empirical fits appear to be satisfactory, especially in the first half of the distributions, where the ``near'' of the term near-extreme actually takes place and it is remarkable how we can fit with a single distribution both near-maximum and near-minimum statistics. In some cases the prediction loses its power when the curve approaches to unity, i.e. when the opposite tail of the log-return distribution begins to play the main role; no longer the ``near''-extreme ones, but actually a ``far''-extreme distribution. As already mentioned, the main aim of this paper is the exploration of a new tool/idea, but nevertheless we can assess the significance of our result performing some simple statistical analyses. We choose to apply the Kolmogorov-Smirnov test (in Tab.~\ref{tab:ks}) and to show the Q-Q plots (in Fig.~\ref{fig:QQplot}). 
The K-S statistics is defined as $D = \max\|P(r^\tau) - P_e(r^\tau)\|$ and the null hypothesis is considered rejected at the $5\%$ significance level, when $\sqrt{\mbox{Sample\,\,size}}\,D$ is larger than $1.333$, and at the $1\%$ significance level, when larger than $1.625$ \cite{Stephens1974}. The only stock presenting systematic problem is INTC, where the fits fail especially when applied to the minimum case. All the other results are satisfactory. For completeness, Tab.~\ref{tab:sample} reports the sample sizes. The Q-Q plots highlight the good agreement of location, scale and skewness of the compared statistics and huge deviations from the straight line are observable only when the quantiles approach unity. In Fig.~\ref{fig:QQplot}, plots for $N=25$ and 50 have been shifted to facilitate readability.

In general, we can say that the method gives satisfactory results for $\tau$ in the range of a few hundreds to roughly ten thousand and for $N$ between 10 to 100. When $\tau$ is too small, the local Gaussianity idea breaks down (given the discrete nature of the prices as shown in Fig.~\ref{fig:minima} and the strong dependencies) and when $N$ is not large enough the estimations of the variances become too noisy. On the other hand, an excessively large value of those parameters (for a large $N$ or $\tau$, $h$ as defined in Sec.~\ref{sec:idea} becomes small) tend to give much more importance to each and every extreme value, and to give a poor statistics of variances. As a final remark we would like to clarify that our ``estimator'' for the distribution as defined in Eq.~\ref{eq:approxmargin1} can be seen as a mixture of the distributions defined in Ref.~\cite{Sabhapandit2007}, exactly in the same fashion used in Ref.~\cite{Gerig2009} to fit the intraday log-return distribution. 

\section*{Acknowledgment}
The stay of MP at ICU has been financed by the Japanese Society for the Promotion of Science (grant N. PE09043).

\bibliographystyle{elsarticle-num}
\bibliography{paper_PhysicaA_FINAL}
\end{document}